\documentstyle[preprint,aps]{revtex}
\begin{document}
\draft
\preprint{}
\title{Yang-Mills Inspired Solutions for General Relativity}
\author{D. Singleton}
\address{Department of Physics, Virginia Commonwealth University, 
Richmond, VA 23284-2000}
\date{\today}
\maketitle
\begin{abstract}
Several exact, cylindrically symmetric solutions to 
Einstein's vacuum equations are given. These solutions
were found using the connection between Yang-Mills
theory and general relativity. Taking known solutions
of the Yang-Mills equations ({\it e.g.} the topological BPS 
monopole solutions) it is possible to construct exact solutions 
to the general relativistic field equations. Although the
general relativistic solutions were found starting from
known solutions of Yang-Mills theory they have different
physical characteristics.
\end{abstract}
\newpage
\narrowtext

\section{The Ernst Equations} 

Recently several exact solutions for the Yang-Mills field 
equations were found \cite{sing} \cite{sing1}
by using the correspondence between Yang-Mills theory
and general relativity. The idea was to use known general
relativistic solutions, such as the Schwarzschild and Kerr
solutions, to find analogous Yang-Mills solutions. It was
conjectured that these solutions may be connected with the
confinement mechanism in QCD, since just as the Schwarzschild
solution of general  relativity will confine any particle
which carries the gravitational ``charge'' behind its
event horizon, so the corressponding Yang-Mills solution
might confine particles behind its ``event horizon''.
Although the Yang-Mills solutions were functionally similiar
to the analogous general relativistic solutions there were
significant physical differences between the two. For example,
the spherical singularity at $r= 2GM$ of the Schwarzschild
solution is a coordinate singularity whereas, in the Yang-Mills 
case the spherical singularity is a true singularity in the gauge
fields and in the energy density. Thus, although the solutions
look functionally similiar, they nevertheless have some
different physical characteristics ({\it e.g.} the $r=2GM$
singularity in the Schwarzschild solution acts as a one way
membrane while the equivalent singularity of the Yang-Mills
solution appears to act as a two way barrier).

In the present paper we want to invert the above process and
use known solutions of the Yang-Mills field equations to
find solutions to Einstein's vacuum equations. The best 
known exact solutions of the Yang-Mills field equations are the
monopole solutions ({\it e.g.} the Prasad-Sommerfield-Bogomolnyi
\cite{som} solution). These Yang-Mills monopole solutions can
be viewed as topological solitons, whose fields are non-singular.
The standard interpretation of this Yang-Mills solution is as
a localized particle which  carries magnetic (and possibly 
electric) charge. In contrast the corresponding general 
relativisitic BPS solutions, which are presented here, do not seem
to have an interpretation as arising from a localized distribution
of gravitational charge (mass-energy). In particular the general
relativisitic version of the BPS monopoles
are not asymptotically flat making their physical meaning
unclear. It is also found that two different forms of the BPS 
monopole solution, give physically different solutions when 
carried over to general relativity.

To examine the connection between solutions to Einstein's 
equations and solutions to the Yang-Mills 
equations we use the Ernst equations \cite{ernst}.
The Ernst equations were originally formulated to
simplify the general relativistic field
equations for axially symmetric solutions (particularly solutions
which were parameterized via the Papapetrou metric). Later
it was shown \cite{fhp} that using the axially symmetric ansatz 
of Manton \cite{manton} one could write the field equations 
of an SU(2) Yang-Mills-Higgs system in the form of the
Ernst equations. For axially symmetric
solutions to Einstein's field equations one can write the
down the metric using the Papapetrou ansatz
\begin{equation}
\label{pap}
ds^2 = f(\rho , z) \Big[ dt - \omega (\rho , z) d \phi \Big] ^2 
- {1 \over f(\rho ,z)} \Big[ e^{2 \gamma (\rho , z)} 
(d \rho ^2 + dz^2) + \rho ^2 d \phi ^2
\Big]
\end{equation}
Plugging this ansatz into the vacuum Einstein field equations
yields four coupled differential equations for the
ansatz functions $f(\rho , z)$ , $\omega (\rho , z)$ and
$\gamma (\rho , z)$ \cite{wald}
\begin{eqnarray}
\label{grfeq}
f \nabla ^2 f &=& \nabla f \cdot  \nabla f - {f^4 \over \rho ^2}
\nabla \omega \cdot \nabla \omega \nonumber \\
\nabla \cdot \Bigg({f^2 \over \rho ^2} & \nabla & \omega \Bigg) = 0
\nonumber \\
{\partial \gamma \over \partial \rho} &=& 
{\rho \over 4 f^2} \left[ \left( {\partial f \over \partial \rho} 
\right) ^2 - \left( {\partial f \over \partial z} \right) ^2 \right]
- {f^2 \over 4 \rho} \left[ \left( 
{\partial \omega \over \partial \rho} \right) ^2 
- \left( {\partial \omega \over \partial z} \right) ^2 \right]
\nonumber \\
{\partial \gamma \over \partial z} &=& {\rho \over 2 f^2} 
{\partial f \over \partial \rho} {\partial f \over \partial z} -
{f^2 \over 2 \rho} {\partial \omega \over \partial \rho} 
{\partial \omega \over \partial z}
\end{eqnarray}
Ernst was able to re-write these field equations through 
the introduction of a complex potential
\begin{equation}
\label{complex}
\epsilon= f + i \Psi
\end{equation}
in terms of which some of the field equations became
\begin{equation}
\label{ernst}
Re  (\epsilon ) \nabla ^2 \epsilon = \nabla \epsilon \cdot \nabla
\epsilon
\end{equation}
or more explicitly
\begin{eqnarray}
\label{ernst1}
f \nabla ^2 f &=& \nabla f \cdot \nabla f - \nabla \Psi \cdot
\nabla \Psi \nonumber \\
f \nabla ^2 \Psi &=& 2 \nabla f \cdot \nabla \Psi
\end{eqnarray}
The function $\omega $ is determined from $\Psi$ via \cite{carmeli}
\begin{equation}
\label{omega}
\nabla \omega = {\rho \over f^2} {\hat {\bf n}} \times \nabla \Psi
\end{equation}
where ${\hat {\bf n}}$ is the unit vector in the azimuthal direction.
Since our ansatz functions only depend on $\rho$ and $z$ we can
re-write the above condition as \cite{islam}
\begin{equation}
\label{omega1}
{\partial \omega \over \partial z}  = - {\rho \over f^2} 
{\partial \Psi \over \partial \rho} \; \; \; \; \;
{\partial \omega \over \partial \rho} = {\rho \over f^2} 
{\partial \Psi \over \partial z}
\end{equation}
Thus if a solution is found to the Ernst equation in terms of the
$f$ and $\Psi$ functions, then one can use Eq. (\ref{omega1})
to determine the original function $\omega$. Then the final
ansatz function $\gamma$ can be determined using the last two
equations of Eq. (\ref{grfeq}). An important point to emphasize
is that once $f$ and $\Psi$ are found (or alternatively $f$ and
$\omega$) then the solution is found up to an intergration,
since the last two equations of Eq. (\ref{grfeq}) automatically
imply that the integrability condition, $\partial ^2 \gamma / \partial z
\partial \rho = \partial ^2 \gamma / \partial \rho \partial z$, is
satisfied. Thus a unique solution to $\gamma$ can 
be given via a line integration \cite{wald}. It may not be possible,
however, to obtain a closed form solution for $\gamma$. 

\section{Monopole Solutions of Einstein's Equations}

Several authors have looked for exact solutions for an 
SU(2) system  using Ernst or modified Ernst equations \cite{fhp}
\cite{chak1}. Using this technique it is possible to construct
monopole and multi-monopole solutions  for the Yang-Mills
field equations. Since the Ernst equation is also used to examine
exact solutions in general relativity, it should be possible
to take the Yang-Mills monopole solutions written in the
formulation of Refs. \cite{fhp} \cite{chak1} and arrive at 
corresponding general relativistic solutions. The BPS monopole solutions
have several good features such as having non-singular fields
and finite energy. It was originally hoped that the general
relativisitic versions would inherit these good features, however,
the general relativistic solutions discussed here either have
metrics whose components are singular at some point or 
they do not become asymptotically flat.

In Refs. \cite{chak1} Chakrabarti and Koukiou obtain various
solutions to the Yang-Mills equations by starting with a seed
solution to a modified version of the Ernst equations. Then by 
applying Harrison-Neugebauer type transformations \cite{har}
\cite{neu} on this seed solution they obtain various monopole 
solutions of the Yang-Mills equations. (Forg{\'a}cs {\it et. al.}
\cite{fhp} work directly with the Ernst equation rather than
the Ernst-like equation used in Ref. \cite{chak1}. However their
version of the BPS monopole solution is much more complicated
than in Ref. \cite{chak1}. Thus it is easier to obtain a
closed form general relativisitic solution starting with the
monopole solution in the form given by Chakrabarti and Koukiou).
Before trying to map over the 
monopole solution into a general relativisitic solution
we will examine the easier example of how the seed solution of
Ref. \cite{chak1} can be used to give  a solution to Einstein's
vacuum equations. The general seed solution used in Ref. \cite{chak1}
is
\begin{equation}
\label{soln1}
f (r) = Exp \Big( {-b r - a r^2 \cos(\theta)} \Big) 
\; \; \; \; \; \Psi (r) = 0
\end{equation}
where $a$ and $b$ are arbitrary constants, and we have used 
spherical coordinates in writting out the solution as in
Ref. \cite{chak1}. This seed solution satisfied a modified 
Ernst equation, which is related to the Ernst equation,
Eq. (\ref{ernst1}), by the transformation $r \rightarrow
1/r$. Thus to turn the solution of Eq. (\ref{soln1}) into a
solution of Eq. (\ref{ernst1}) we apply the same transformation to
the solution. This yields
\begin{equation}
\label{soln1a}
f(\rho ,z) = Exp \left( -{b \over \sqrt{\rho ^2 + z^2}} - {a z \over
(\rho ^2 + z^2) ^{3/2}} \right) \; \; \; \; \; \Psi (\rho ,z) = 0
\end{equation}
where we have also changed from spherical to cylindrical
coordinates, since these are the coordinates in which the
original ansatz functions in Eq. (\ref{grfeq}) are formulated. 
It is easily checked by direct substitution
that Eq. (\ref{soln1a}) solves Eq. (\ref{ernst1}). Since
$\Psi (\rho ,z) =0$ in the above solution $\omega (\rho , z) =0$
by Eq. (\ref{omega1}). This is a static solution
with no angular momentum. Using $f$ from Eq. (\ref{soln1a})
the function $\gamma (\rho , z)$ can be found by integrating the 
last two equations of Eq. (\ref{grfeq}) to yield a closed form result
\begin{equation}
\label{gsoln1}
\gamma (\rho ,z) = {9 a^2 \rho ^4 \over 16 (\rho ^2 +z^2)^4}
- {a \rho ^2 (a + b z) \over 2 (\rho ^2 + z^2)^3} -
{b^2 \rho ^2 \over 8 (\rho ^2 +z^2)^2}
\end{equation}
It is easy to see that this solution becomes asymptotically
flat ({\it i.e.} $f \rightarrow 1$, $\omega = \gamma \rightarrow
0$ as $\sqrt{\rho ^2 + z^2} = r \rightarrow \infty$). By looking
at the far field behaviour of this solution we find that the
Newtonian potential is
\begin{equation}
\label{newton}
\Phi = (g_{00} - 1) / 2 = (f -1) /2 \approx -{b \over 2 r} +
{1 \over 2 r^2} \left( {b^2 \over 2} - a \cos(\theta ) \right)
\end{equation}
Thus the constant $b$ is related to the mass of the solution
({\it i.e.} $b = 2GM$) and the $a$ term looks like a dipole
term. If $a =0$ we just recover the Curzon metric \cite{curzon}.
If $b =0$ the Newtonian far field 
potential looks like the dipole potential of
electromagnetism. This could be taken to indicate that this
special case of the solution is not physical. However recent work
\cite{tom} on massless black holes also finds a Newtonian
potential whose leading term falls off like $1/ r^2$ rather
than $1 / r$. The physical interpretation 
of these massless objects was
as bound states of positive and negative mass. For the $b=0$ case
of the above solution it may be even more appropriate to consider
the possibility that the solution represents some kind of 
positive-negative mass bound state since the far field has exactly the
kind of angular dependence one would expect of a dipole field, while
the solution in Ref. \cite{tom} only has the $1/r^2$ behaviour,
but not the dipole angular dependence. In a certain
sense the solution given by Eqs. (\ref{soln1a}) and (\ref{gsoln1})
is not mathematically very interesting since it is just a
specific example of a Weyl solution. However it is an
asymptotically free, closed form solution, and in light of
Ref. \cite{tom} it may be of some physical interest. Although
this solution becomes asymptotically flat it has the undesirable
feature that some of the components of its metric become singular 
at $r=0$ ({\it e.g.} $g_{3 3}$ diverges as $\sqrt{ \rho^2 +z^2}
= r \rightarrow 0$ since $f \rightarrow 0$).

A more interesting solution, which is not just a particular example
of a Weyl solution is the BPS monopole solution. In Ref. \cite{chak1}
it is found that the BPS monopole can be expressed in terms of the
ansatz functions of the Ernst-like equation as
\begin{equation}
\label{soln2}
f(r) = csch(r) \; \; \; \; \; \Psi (r) = i \; \coth (r)
\end{equation}
Where we have again written the solution in spherical coordinates.
In this form the connection to the BPS solution is very apparent
since these are exactly the hyperbolic functions used
in the BPS monopole solution.
Since $\Psi (r)$ is imaginary, $\omega$ will also
be imaginary, which makes the solution unphysical. However taking
the complementary hyperbolic functions of the above solution we
find that we can get a completely real valued solution. Making the
transformation $r \rightarrow 1/r$ (so that our solution satisfies
the Ernst equation rather than the modified Ernst equation of Ref. 
\cite{chak1}), and switching to cylindrical coordinates (so that
the solutions can be checked in Eqs. (\ref{grfeq})) we find the 
following real solution to the Ernst equation
\begin{equation}
\label{soln2a}
f(\rho , z) = D \; sech \left( a + {b \over \sqrt{\rho ^2 + z^2}} \right)
\; \; \; \; \; \;
\Psi (\rho , z) = D \; \tanh \left( a + {b \over \sqrt{\rho ^2 +z^2}}
\right)
\end{equation}
where $D$, $a$, and $b$ are constants, and we have generalized
the solution somewhat by introducing the constant $a$.
Using Eq. (\ref{omega1}) we can determine the ansatz function
$\omega$ from $\Psi$. Integrating the equations gives
\begin{equation}
\label{omsoln2}
\omega (\rho , z) = {b z \over D \sqrt{\rho ^2 + z^2}} = 
{b  \cos (\theta ) \over D}
\end{equation}
where in the last step we have written the result in spherical
coordinates. Since $\omega \ne 0$ this is a stationary solution 
as opposed to the first solution which was static. A non-zero
$\omega$ usually indicates a source with some angular momentum.
However, for a body with angular momentum $S$ one would expect
$\omega \rightarrow -2 S \rho ^2 / (\rho^2 +z^2) ^{3/2}$ as
$r = \sqrt{ \rho ^2 +z^2} \rightarrow \infty$ \cite{islam} 
which is not
the case here. This behaviour of $\omega$ indicates that
this solution does not become asymptotically flat, and makes
a physical interpretation difficult. There are other known
stationary solutions, such as the NUT-Taub metric \cite{taub}, 
the Lewis metric \cite{lew} and the Van Stockum metric \cite{van},
which also do not become asymptotically flat. The ansatz
function $\omega$ can be made small by letting
$D$ become large and/or allowing $b$ to become small.
Finally using Eq. (\ref{grfeq}) we can determine
the last ansatz function
\begin{equation}
\label{gsoln2}
\gamma (\rho , z) = {-b^2 \rho ^2 \over 8 (\rho ^2 +z^2)^2} = 
{-b^2 \sin ^2 (\theta ) \over 8 r^2}
\end{equation}
where in the last step we have again used spherical
coordinates. Thus this form of the BPS monopole solution gives 
an exact, closed form solution to Einstein's vacuum equations, 
and the components of the metric tensor are non-singular except
at the origin (in particular $g_{33} \rightarrow \infty$ as $r
\rightarrow 0$). At large distances ({\it i.e.} $r \rightarrow
\infty$) $\gamma \rightarrow 0$ as one would want for an
asymptotically flat solution. Also if one requires that
$D = \cosh(a)$ then $f \rightarrow 1$ as $r \rightarrow \infty$
also as expected for an asymptotically flat solution. However
$\omega$ does not go to zero at large distances (except in the $x-y$
plane where  $z=0$ or $\theta = \pi / 2$), thus this solution
can not be viewed as some finite, localized distribution of
rotating mass. To determine the physical meaning of the arbitrary 
constants of this solution we can again examine the Newtonian 
potential at large distances
\begin{eqnarray}
\label{newton1}
\Phi &=& (f -1)/2 \approx  {-1 + D sech (a) \over 2} -
{bD sech (a) \tanh (a) \over 2 r} \nonumber \\
&+& {D sech (a) \big( -b^2/2 + b^2 \tanh ^2 (a) \big) \over 2 r^2} +
{\cal O} (1/r^3) 
\end{eqnarray}
First we can choose $D = \cosh (a)$ so that the leading term of
the potential goes as $1/r$. Then we can set $b \tanh (a) =
2GM$ so that $a$ and $b$ appear to be related to the mass of the 
solution. Finally as a special case we could take $\tanh (a) =0$.
In this case we would obtain a Newtonian potential similiar to
that of the massless Reissner solution or to the far field
found in Ref. \cite{tom}. It is not clear what 
physical use if any this closed form solution to the 
vacuum equations may have. Since this solution is not
asymptotically flat, it can not represent the exterior field
of some localized distribution of rotating matter. Since only
$\omega$ does not approach its correct asymptotic value, this 
solution appears to have a source with an infinite angular
momentum. The similiarity between this solution and the NUT-Taub
solution \cite{taub} should be pointed out. The Newtonian potential
of both the present metric and the NUT-Taub metric fall off as
$1/r$ at large distances. More importantly the $g_{03}$ term
of both solutions have exactly the same form, and this keeps
both solutions from being asymptotically flat.

Finally it was shown in Ref. \cite{chak1} that the BPS
monopole could also be obtained from the following alternative
form of the solution to the modified Ernst equations
\begin{equation}
\label{soln3}
f (\rho , z) = {\sinh (r) \sin(\theta ) \over r} 
\; \; \; \; \; \Psi (\rho , z) = \cos(\theta )
\end{equation}
Going through the usually transformation to turn this into a
solution of the regular Ernst equation ({\it i.e.} $r \rightarrow
1/r$), scaling the variable $r$ ({\it i.e.} $r \rightarrow b r$),
and finally converting to cylindrical coordinates we arrive at
the follwoing solution to the Ernst equations
\begin{equation}
\label{soln3a}
f (\rho , z) = b \; \rho \sinh \left( {1 \over b \sqrt{\rho ^2 + z^2}} \right)
\; \; \; \; \;  \Psi (\rho , z) = {z \over \sqrt{\rho ^2 + z^2}}
\end{equation}
Using the above function $\Psi$ in Eq. (\ref{omega1}) we find the
ansatz function $\omega$
\begin{equation}
\label{om5}
\omega (\rho , z) = {1 \over b} \coth 
\left( {1 \over b \sqrt{ \rho ^2 +z^2}} \right)
\end{equation}
Already one can see that this solution will not give an asymptotically
flat solution, since as $\sqrt{ \rho ^2 +z^2} = r \rightarrow
\infty $,  $\omega \rightarrow \infty$ rather than $0$. The last
ansatz function, $\gamma$, can not be obtained in closed form
in this case. As $r \rightarrow \infty$ one can obtain the 
following approximate form for the equations that determine $\gamma$
\begin{eqnarray}
\label{gsoln3}
{\partial \gamma \over \partial \rho} 
\approx {z^2 - \rho ^2 \over 4 \rho (\rho ^2 + z^2)}
\nonumber \\
{\partial \gamma \over \partial z} \approx {-z \over 2 (\rho ^2 + z^2)}
\end{eqnarray}
These can be integrated to give
\begin{equation}
\label{gsolna}
\gamma (\rho , z) \approx {1 \over 4} \ln 
\left( {\rho \over \rho ^2 + z^2} \right)
\end{equation}
which is valid as $\sqrt{\rho ^2 + z^2} = r \rightarrow \infty$.
Thus the ansatz function, $\gamma$, does not indicate an asymptotically
flat solution. The asymptotic behaviour here is worse than for
the previous solution since both $\omega$ and $\gamma$
diverge as $r \rightarrow \infty$.  In addition $f$ is 
divergent as $r \rightarrow 0$. Even though the solutions 
of Eq. (\ref{soln2}) and Eq. (\ref{soln3}) yield
the same field configuration for the Yang-Mills equations ({\it i.e.}
they both give the BPS monopole) they give apparently different
solutions when carried over to general relativity. Also
the physical characteristics of the Yang-Mills solution do not
necessarily carry over into the general relativisitic solution
({\it e.g.} the BPS monopole is well behaved  over all space, while
the three general relativisitic solutions presented here have 
some undesired features : they are not asymptotically flat or the
components of the metric become singular at certain points). At this point 
it is not known whether the singularities are real features of
these solutions or whether they might not be coordinate singularities
as is the case for the event horizon of the Schwarzschild solution.

\section{Discussion and Conclusions}

We have presented three solutions to Einstein's vacuum field
equations which were found by exploiting the connection between
Yang-Mills theory and general relativity. In Ref. \cite{fhp}
it was shown that the Yang-Mills field equations could be put
in the form of the Ernst equations of general relativity through
the use of Manton's ansatz \cite{manton}. Previously this
framework has been used by the author map the Kerr solution
of general relativity into a Yang-Mills counterpart
\cite{sing1}. Here we have carried out this procedure 
in reverse : by starting with the known monopole solutions 
to the Yang-Mills field equations we obtained new solutions 
to the general relativistic field equations.
In doing this we used the BPS monopole solution in the forms
given in Ref. \cite{chak1} where the monopole solution 
was derived from a modified Ernst equation. This made 
it straight forward to covert the Yang-Mills solutions to
general relativistic counterparts. The first solution studied
in this paper was the general seed solution used
in Ref. \cite{chak1} to obtain the monopole solutions via
Harrison-Neugebauer tranformations. This Yang-Mills solution
gave a general relativistic solution, which was a generalization
of the Curzon metric. In the special case of this solution where
the constant $b$ was set to zero ({\it i.e.} the mass of the 
solution became zero) we found that the far field Newtonian 
potential behaved like a dipole field. Thus this solution may
have some connection with some recent work on black diholes
\cite{tom}. This solution had a singularity at the origin in
some of the components of its metric and
it became asymptotically flat. The second solution which we 
examined -  Eq. (\ref{soln2}) - 
was one form of the BPS monopole. The general relativistic
version of this solution - Eqs. (\ref{soln2a}) (\ref{omsoln2}) 
(\ref{gsoln2}) - did not become asymptotically
flat as $r \rightarrow \infty$ (in particular $\omega$ did not
go to zero). This may be related to the fact that in the
BPS solution the Higgs field does not go to zero as $r \rightarrow 
\infty$. Since $\omega$ did not go to zero asymptotically
one could interpret this solution as having a source with infinite
angular momentum. The final solution which we examined was
a different version of the BPS monopole. This alternative form of
the BPS monopole gave a general relativisitic solution with a
different asymptotic behaviour and different physical 
characterisitics from the general relativisitic solution
obtained from the version of the BPS monopole given by Eq.
(\ref{soln2}). This high lights the fact that although solutions 
of one theory can be used to find solutions in the other,
the physical characteristics of the original solution are not
all necessarily inherited by the new solution. For example,
in the Yang-Mills version of the Schwarzschild solution
the spherical singularity of the solution is a true singularity,
while for the general relativisitic Schwarzschild solution the
spherical singularity is a coordinate singularity. Nevertheless,
both Yang-Mills and general relativity do seem to share some
degree of mathematical similarity at the level of the classical
field equations, which allows one to use the solutions of
one theory to obtain solutions in the other.

\section{Acknowledgements} The author would like to thank Tom
Michael and Prof. Bob Jones of MIT for interesting discussions
and encouragement.

\end{document}